\documentclass[aps,prb,twocolumn,groupedaddress,showpacs,floatfix,altaffilletter,longbibliography]{revtex4-1}

\usepackage{amsfonts}
\usepackage{amsmath}
\usepackage{amssymb}
\usepackage{graphicx}% Include figure files
\usepackage{mathrsfs}
\usepackage{mathtools}
\usepackage{grffile}
\usepackage{amsmath}
\usepackage{amssymb}
\usepackage{bm}
\usepackage{color}
\usepackage{amsmath,amssymb,amsfonts}
\usepackage{graphicx}% Include figure files
\usepackage{times}
\usepackage[pdftex]{hyperref}
\usepackage[pdftex]{hyperref}
\hypersetup{plainpages=false,colorlinks=true,linkcolor=red, citecolor=blue, urlcolor=blue}%

\newcommand{\fref}[1]{Fig.~\ref{#1}}
\newcommand{\eref}[1]{Eq.~(\ref{#1})}

\begin{document}
%\definecolor{qqqqff}{rgb}{0.,0.,1.}
%\definecolor{qqwuqq}{rgb}{0.6,0.56,0.76}
%\definecolor{ffqqqq}{rgb}{0.9,0.38,0.}
%\definecolor{zzttqq}{rgb}{0.6,0.2,0.}

%\title{Symmetry properties of correlation functions in orbital basis:  the role of the nonsymmorphic space group in spin density wave phase of iron-based superconductors }

\title{Effect of non-symmorphic space groups on correlation functions in iron-based superconductors }

\author{R.~Nourafkan$^{1}$} 
%\email{Reza.Nourafkan@usherbrooke.ca}
\author{A.-M.S. Tremblay$^{1,2}$}
%\email{Andre-Marie.Tremblay@usherbrooke.ca }
\affiliation{$^1$D{\'e}partement de Physique and Regroupement qu\'eb\'ecois sur les matériaux de pointe, Universit{\'e} de Sherbrooke, Sherbrooke, Qu{\'e}bec, Canada J1K 2R1}
\affiliation{$^2$Quantum Materials Program, Canadian Institute for Advanced Research, Toronto, Ontario,  M5G 1Z8, Canada}
%
%\date{\today}
%
\begin{abstract}
The orbital basis is natural when one needs to calculate the effect of local interactions or to unravel the role of orbital physics in the response to external probes. In systems with nonsymmorphic point groups, such as the iron-based superconductors, we show that symmetries that emerge in observable response functions at certain wave vectors are absent from generalized susceptibilities calculated with tight-binding Hamiltonians in the orbital basis. Such symmetries are recovered only when the generalized susceptibilities are embeded back to the continuum using appropriate matrix elements between basis states. This is illustrated with the case of LiFeAs and is further clarified using a minimal tight-binding Hamiltonian with non-symmorphic space group. 
\end{abstract}
%
%\pacs{74.20.Pq, 74.70.Xa}
% 74.20.Pq	Electronic structure calculations
%74.70.Xa	Pnictides and chalcogenides

\maketitle
\section{Introduction}

A density or magnetic response function measures the response to an applied conjugate field. The density response function, for example, is given by $\langle \hat{n}({\bf r},t)\hat{n}({\bf r}',0) \rangle$ where $\hat{n}$ denotes the density operator.  
Interactions between electrons cause correlations that complicate the calculation of response functions: The electron and the hole created by an external field interact not only separately with the other particles but also with each other.  
%There are different approximate methods to account for correlation effects given by $\langle \hat{n}({\bf r},t)\hat{n}({\bf r}',0) \rangle - \langle \hat{n}({\bf r},t)\rangle \langle\hat{n}({\bf r}',0) \rangle$.  
Since correlations affect different orbitals differently, developing a microscopic understanding requires expressing the response function in a suitable orbital basis and studying its different components.  
How these different components are transformed under point-group symmetry operations is one of the questions that arises naturally.

The symmetry properties of the observable response function is usually discussed in the  long wave-length (small wave-vector) limit where the inner structure of the unit cell becomes irrelevant.~\cite{Bassani1975}  However, a large number of recent experiments are done at finite wave-vectors that probe the inner structure of the unit cell.
Furthermore, all approximations for interaction effects are formulated in an atomic-like orbital basis where interaction strengths can be defined more straightforwardly than in the extended Bloch states.  
Therefore, it is important to verify how the underlying lattice influences the symmetry properties of the correlation functions calculated in the orbital basis. %This analysis is also helpful to distinguish the possible perturbations which break a given symmetry.
 
One of our main messages is that for non-symmorphic crystals it is necessary to map back the model to the continuum to recover all the symmetries of observable susceptibilities. The generalized susceptibilities in orbital space have a lower symmetry. That result is very general. But we find it preferable to include a specific example. We chose the iron-based superconductors (FeSCs) which are still actively studied. (For a review see Ref.~\onlinecite{HOSONO2015399}). Similar to the copper oxide high temperature superconductors, these are  layered materials where the superconductivity arises in proximity to long-range magnetic order. However, the nature of the magnetic orders differs between the two families. In FeSCs the magnetic order is a stripe-type antiferromagnetic (AF) which lowers the $S_4$ symmetry (the fourfold rotational symmetry, $C_4$, around an Fe-site followed by reflection about the plane) to $C_2$. This phase is necessarily accompanied by a structural transition (tetragonal-to-orthorhombic).

The FeSCs crystalize in a non-symmrophic space group.
Numerous studies have investigated the impacts of the non-symmorphic space group of FeSCs on their electronic properties in the normal  or superconducting phases.~\cite{PhysRevB.82.134503, Hirschfeld2016197, PhysRevB.78.144517, PhysRevB.88.134510, PhysRevLett.114.107002, PhysRevX.4.031053, PhysRevX.3.031004, PhysRevB.92.214509, 1367-2630-15-7-073006} They either discuss the symmetry of the Bloch states at high symmetry points of the reciprocal lattice or consider possible unconventional Cooper pairing. We instead consider explicit symmetries of the generalized susceptibilities and of the observable susceptibilities in the orbital basis which, to our knowledge, have not been discussed so far.  Our goal in this paper is to present a detailed discussion of these symmetries.

One of the non-symmorphic symmetry operations of  the FeSCs is a $\pi/2$ rotation about an axis perpendicular to the iron-pnictide planes (and centered in the middle of the bond separating two irons) followed by a half-integer translation (see Fig.~\ref{fig1}). 
%A model Hamiltonian in the orbital basis is invariant under that symmetry operation. 
We show in this paper that the above non-symmorphic symmetry operation leads, as expected, to generalized correlation functions that are {\it not} invariant under $\pi/2$ rotation alone when calculated with a model Hamiltonian in the orbital basis since rotation by $\pi/2$ alone is not a symmetry. There is however an apparent paradox since correlation functions evaluated at wave vector $q_z=0$ should be invariant under $\pi/2$ rotation, as follows from the fact that rotation by $\pi/2$ around an iron atom in the $xy$ plane, followed by a reflection $z \rightarrow -z$  leaves the lattice invariant. Here we solve this paradox by showing that to recover the $\pi/2$ symmetry at $q_z=0$, one must take into account overlap matrix elements between local basis states. These are necessary to embed the orbital-basis model back to the continuum. We call these matrix elements oscillator matrix elements.  Note that similar considerations apply to lattices whose space-group is symmorphic when a non-primitive cell is chosen. For these lattices however, a primitive cell can always be chosen so that the unavoidable complications that we face in the non-symmorphic case can be avoided in the symmorphic case. 
 
The rest of this paper is organized as follows.  In  section \ref{Hamiltonian} we recall the procedure to obtain the Hamiltonian in the orbital basis. Sec. \ref{Susceptibility} introduces the generalized charge susceptibilities that are usually calculated in the orbital basis and their link to observable charge susceptibilities in the continuum through embedding with oscillator matrix elements. 
Then we  explain in section \ref{Transformation} how the orbital-basis Hamiltonian transforms under non-symmorphic operations. The discussions in these sections  are general. Sect.~\ref{Result} presents some results on the bare susceptibility of LiFeAs. Sect.~\ref{Con} contains  concluding remarks. Appendix \ref{AppC} demonstrates how oscillator matrix elements appear in the two-body interaction in the orbital-basis, appendix \ref{AppA} gives group-symmetry details of LiFeAs, and  appendix \ref{AppB} explicitly shows how extra phases appear in the orbital basis because of non-symmorphic transformations. These transformations are further explained in appendix \ref{TBHamiltonian} using a minimal tight-binding Hamiltonian to explicitly show that even though $q_z=0$ eigenenergies are invariant under $±\pi/2$ rotation, eigenvectors are not. We also show explicitly in this appendix how oscillator matrix elements allow us to recover $q_z=0$ symmetry of the charge fluctuations.
 
\section{Hamiltonian in the orbital basis}\label{Hamiltonian}
In this section we recall how to write the Hamiltonian in the orbital basis.  This allows us to introduce the notation. We emphasize that the discussions in this section and the following section are general. The non-interacting part of the Hamiltonian is given by
\begin{align}
\hat{\bf {\mathcal H}}= \sum_{\sigma}\int d{\bf r} 
\hat{\Psi}^{\dagger}_{\sigma}({\bf r}) [-\frac{\hbar^2}{2m}\nabla^2 +\sum_{{\bf R}_i}V({\bf r}-{\bf R}_i)]
\hat{\Psi}_{\sigma}({\bf r}) ,\label{Ham}
\end{align}
where $\hat{\Psi}_{\sigma}({\bf r},\tau)$ is the field operator, $V({\bf r}-{\bf R}_i)$ denotes the ionic potential and ${\bf R}_i$ denotes the lattice sites. The field operators  can be expanded in a basis set.  In a weakly correlated crystalline system, the prime candidate for the basis set is Bloch functions. However, in an interacting system it is more suitable to expand the field operators in terms of an atomic-like orbital basis set $\{\phi_{l{\bf R}_i}({\bf r})\}$, namely, 
\begin{equation}
\hat{\Psi}_{\sigma}({\bf r}) = \sum_{l{\bf R}_i} \phi_{l{\bf R}_i,\sigma}({\bf r}) \hat{c}_{l{\bf R}_i,\sigma}
\label{Psi}
\end{equation}
where $\hat{c}$ denotes the second quantized destruction operator for the state described by the atomic position ${\bf R}_i$ and by the index $l\equiv(\alpha,l^{\alpha})$ that combines ion (sublattice), $\alpha$, and orbital indices $l^{\alpha}$.  We assume that the basis set is an orthonormal Wannier basis with zero direct overlap between different lattice sites, 
\begin{align}
\int d{\bf r} \phi^*_{ l_1{\bf R}_1,\sigma}({\bf r}) \phi_{ l_2{\bf R}_2,\sigma'}({\bf r}) = \delta_{\sigma,\sigma'}\delta_{l_1l_2}\delta_{{\bf R}_1,{\bf R}_2}.
\end{align}
Therefore, the single-particle Hamiltonian can be written as
\begin{align}
\hat{\bf {\mathcal H}}=\sum_{\sigma} &\sum_{l_1 l_2} \sum_{{\bf R}_1,{\bf R}_2}{\bf H}^{l_1l_2,\sigma}_{{\bf R}_1,{\bf R}_2}\hat{c}^{\dagger}_{ l_1{\bf R}_1,\sigma}\hat{c}_{ l_2{\bf R}_2,\sigma} ,\nonumber \\
{\bf H}^{l_1l_2,\sigma}_{{\bf R}_1,{\bf R}_2}&=\int d{\bf r}  \phi^*_{ l_1{\bf R}_1,\sigma}({\bf r})[-\frac{\hbar^2}{2m}\nabla^2 \nonumber \\ &+\sum_{{\bf R}_i}V({\bf r}-{\bf R}_i)] \phi_{ l_2{\bf R}_2,\sigma}({\bf r}) .\label{Ham0}
\end{align}

The two commonly used gauges for the periodic orbital basis are the standard gauge where  
\begin{align}
\phi_{l{\bf k},\sigma}=\frac{1}{\sqrt{N}}\sum_{{\bf R}_i} \exp(i{\bf k}\cdot {\bf R}_i) \phi_{l{\bf R}_i,\sigma}\label{phik}
\end{align}
with $N$ the number of unit cells, and the alternative gauge where 
\begin{align}
\phi_{\alpha l^{\alpha}{\bf k},\sigma}=\frac{1}{\sqrt{N}}\sum_{{\bf R}_i} \exp[i{\bf k}\cdot ({\bf R}_i+{\bf r}_{\alpha})]\phi_{\alpha l^{\alpha}{\bf R}_i,\sigma}
\end{align}
 with ${\bf r}_{\alpha}$ the position of atom $\alpha$ within a unit cell. 
 %This alternative gauge is useful when one couples the electromagnetic field through the Peierls substitution. 
 In the standard gauge, the basis functions are strictly periodic in the Brillouin zone (BZ), hence ${\bf H}_{\bf k}$ is BZ periodic, i.e, ${\bf H}_{{\bf k}+{\bf G}} = {\bf H}_{\bf k}$ where ${\bf G}$ is a primitive reciprocal lattice vector. In the alternative gauge, the  Hamiltonian is given by ${\bf {\mathcal H}}=\sum_{{\bf k}}{\bm \phi}^{\prime \dagger}_{{\bf k}}{\bf H}^{\prime}_{\bf k} {\bm \phi}^{\prime}_{{\bf k}}$ where ${\bm \phi}^{\prime}_{\bf k}={\bf V}_{\bf k}^\dagger{\bm \phi}_{\bf k}$ and ${\bf H}^{\prime}_{\bf k}={\bf V}^{\dagger}_{\bf k} {\bf H}_{\bf k} {\bf V}_{\bf k}$ where ${\bf V}_{\bf k}$ is a diagonal unitary matrix with diagonal elements $\exp(-i{\bf k}\cdot {\bf r}_{\alpha})$ and ${\bf H}_{\bf k}$ is Hamiltonian matrix in the standard gauge. In this gauge,  ${\bf H}^{\prime}_{\bf k}$ is not periodic and obeys the relation, ${\bf H}^{\prime}_{{\bf k}+{\bf G}} = {\bf V}^{\dagger}_{\bf G} {\bf H}^{\prime}_{\bf k} {\bf V}_{\bf G}$, which should be accounted for whenever the eigenstates of  the Hamiltonian are needed on the boundary of BZ or outside of it, as occurs in calculations of correlation functions.  For this reason we work in the standard gauge here. However, we will also comment on the alternative gauge.

By introducing the relations
\begin{align}
\hat{c}_{l{\bf k},\sigma} &= \frac{1}{\sqrt{N}}\sum_{{\bf R}_i} \exp(-i{\bf k}\cdot {\bf R}_i) \hat{c}_{l{\bf R}_i,\sigma},\label{FT}\\
\hat{c}_{l{\bf R}_i,\sigma} &= \frac{1}{\sqrt{N}}\sum_{{\bf k}} \exp(i{\bf k}\cdot {\bf R}_i) \hat{c}_{l{\bf k},\sigma},\label{IFT}
\end{align}
and substituting them in the Hamiltonian~\eref{Ham0}, one obtains
\begin{align}
\hat{\bf {\mathcal H}}&=\sum_{{\bf k},\sigma} \sum_{l_1 l_2} {\bf H}^{l_1l_2}_{{\bf k}\sigma} \hat{c}^{\dagger}_{ l_1{\bf k},\sigma}\hat{c}_{ l_2{\bf k},\sigma} ,\nonumber \\
{\bf H}^{l_1l_2}_{{\bf k}\sigma}&=\sum_{{\bf R}_i}e^{-i{\bf k}\cdot {\bf R}_i} {\bf H}^{l_1l_2,\sigma}_{{\bf R}_i,{\bm 0}}
,\label{Ham1}
\end{align}
where we used the fact that ${\bf H}^{l_1l_2,\sigma}_{{\bf R}_1,{\bf R}_2}$ depends on the relative distance between sites ${\bf R}_1$ and ${\bf R}_2$, or ${\bf H}^{l_1l_2,\sigma}_{{\bf R}_1+{\bf R}_0,{\bf R}_2+{\bf R}_0} = {\bf H}^{l_1l_2,\sigma}_{{\bf R}_1,{\bf R}_2}$, and the fact that the ionic potential  is periodic $\sum_{{\bf R}_i}V({\bf r}+{\bf R}_0-{\bf R}_i) = \sum_{{\bf R}_j}V({\bf r}-{\bf R}_j)$.
\eref{Ham1} is what we call the Hamiltonian in the orbital basis.

In matrix form, the Hamiltonian in the orbital basis is given by ${\bf {\mathcal H}}=\sum_{{\bf k}\sigma} {\bm \phi}^{\dagger}_{{\bf k}\sigma}{\bf H}_{{\bf k}\sigma} {\bm \phi}_{{\bf k}\sigma}$,
where ${\bm \phi}_{\bf k} \equiv (c_{1,{\bf k}\sigma},\ldots, c_{n_{orb},{\bf k}\sigma})^T$ and $n_{orb}$ denotes the number of orbitals within a unit cell (This counts all atoms and their respective orbitals). It is worth recalling that while both ${\bm \phi}_{{\bf k}\sigma}$ and ${\bf H}_{{\bf k}\sigma}$ are basis and gauge dependent, the Hamiltonian ${\hat {\mathcal H}}$ is an observable and is gauge invariant.

\section{Charge susceptibility}\label{Susceptibility}
\subsection{Definitions and link between susceptibility of an orbital basis model and observable susceptibility}
In this section, we introduce a generalized susceptibility Eq.~\eqref{Chi0} of the type encountered in orbital-basis calculations and its relation to the susceptibility observable in the continuum, Eq.~\eqref{ChargeSus}. This section then also defines oscillator matrix elements Eq.~\eqref{OSReal} and \eref{OM} that are necessary to connect the two. 

To be specific, consider the density-density response function given by
\begin{align}
\chi({\bf r},{\bf r}',\tau) = 
\langle T_{\tau}\hat{n}({\bf r},\tau) \hat{n}({\bf r}^{\prime} ,0) \rangle_c,
\end{align}
where  the density operator can be written as $\hat{n}({\bf r},\tau) = \sum_{\sigma}\hat{\Psi}^{\dagger}_{\sigma}({\bf r},\tau)\hat{\Psi}_{\sigma}({\bf r},\tau)$. The subscript $c$ on the average reminds us that we take only the connected part. The operators here evolve in Matsubara imaginary time $\tau$. In Fourier space, using the relation $\chi({\bf r},{\bf r}',\tau)=(1/N)\sum_{{\bf R}_0}\chi({\bf r}+{\bf R}_0,{\bf r}'+{\bf R}_0,\tau)$ where ${\bf R}_0$ is a lattice vector,  we have
\begin{align}
\chi_{{\bf G}{\bf G}^{\prime}}({\bf q},\tau) = \frac{1}{V}\int d{\bf r} d{\bf r}' e^{-i({\bf q}+{\bf G})\cdot{\bf r}} \chi({\bf r},{\bf r}',\tau)e^{i({\bf q}+{\bf G}^{\prime})\cdot{\bf r}'},
\end{align}
where the integral is over the entire volume $V$. Here, ${\bf q}+{\bf G}^{\prime}$ and ${\bf q}+{\bf G}$ denote respectively the external field and the response wave-vectors which may fall outside of the BZ. 
Similarly by using the field operator expansion Eq.~\eqref{Psi} in terms of the atomic-like orbitals we find

\begin{align}
\chi_{{\bf G}{\bf G}^{\prime}}&({\bf q},\tau) = \sum_{\sigma \sigma'} \sum_{l_1\ldots l_4}\sum_{{\bf R}_1\ldots {\bf R}_4,}
{\bf O}^{l_1l_2,\sigma}_{{\bf R}_1{\bf R}_2} ({\bf q}+{\bf G}) \nonumber \\ &\times
\langle T_{\tau}\hat{c}^{\dagger}_{ l_1{\bf R}_1,\sigma}(\tau)\hat{c}_{ l_2{\bf R}_2,\sigma}(\tau)  
\hat{c}^{\dagger}_{ l_3{\bf R}_3,\sigma'}(0)\hat{c}_{ l_4{\bf R}_4,\sigma'}(0) \rangle_c\nonumber \\ &\times
{\bf O}^{l_3l_4,\sigma'}_{{\bf R}_3{\bf R}_4} (-{\bf q}-{\bf G}^{\prime}),\label{Resp000}
\end{align}
where we defined the oscillator matrix element as
\begin{align}
{\bf O}^{l_1l_2,\sigma}_{{\bf R}_1{\bf R}_2} ({\bf q})&\equiv\frac{1}{\sqrt{V}}\int d{\bf r} 
e^{-i{\bf q}\cdot {\bf r}} 
\phi^*_{{\bf R}_1 l_1,\sigma}({\bf r}) \phi_{{\bf R}_2 l_2,\sigma}({\bf r}).\label{OSReal}
\end{align}
Using the relation,  $\phi_{{\bf R}_1+{\bf R}_0 l_1,\sigma}({\bf r})=\phi_{{\bf R}_1 l_1,\sigma}({\bf r}-{\bf R}_0)$, one finds that the oscillator matrix elements satisfy 
\begin{align}
{\bf O}^{l_1l_2,\sigma}_{{\bf R}_1+{\bf R}_0,{\bf R}_2+{\bf R}_0}({\bf q}) = \exp(-i{\bf q}\cdot {\bf R}_0){\bf O}^{l_1l_2,\sigma}_{{\bf R}_1,{\bf R}_2}({\bf q}).
\end{align}
Substituting the Fourier-space basis for the second-quantized operators, \eref{IFT}, in \eref{Resp000} for the susceptibility, one finds 
\begin{align}
\chi_{{\bf G}{\bf G}^{\prime}}({\bf q}&,\tau) = \sum_{\sigma\sigma'}\sum_{{\bf k}{\bf k}'}\sum_{l_1\ldots l_4}
{\bf O}^{l_1l_2,\sigma}_{{\bf k},{\bf k+q}} ({\bf q}+{\bf G})
\nonumber \\
\times&\langle T_{\tau}\hat{c}^{\dagger}_{{\bf k} l_1,\sigma}(\tau)\hat{c}_{{\bf k+q} l_2,\sigma}(\tau)
\hat{c}^{\dagger}_{{\bf k}' l_3,\sigma}(0)\hat{c}_{{\bf k'-q} l_4,\sigma} (0)\rangle_c\nonumber \\
\times&{\bf O}^{l_3l_4,\sigma'}_{{\bf k}',{\bf k}'-{\bf q}} (-{\bf q}-{\bf G}^{\prime})
,\label{Resp0}
\end{align}
where it is understood that ${\bf k}+{\bf q}$ and ${\bf k'}-{\bf q}$ must be folded back to the first Brillouin zone with the appropriate Umklapp reciprocal lattice vectors ${\bf G}_0$ and ${\bf G}_0^{\prime}$, as follows from the following identity
\begin{align}
&\sum_{{\bf R}_1{\bf R}_2} e^{-i{\bf k}_1\cdot {\bf R}_1} e^{i{\bf k}_2\cdot {\bf R}_2}{\bf O}^{l_1l_2,\sigma}_{{\bf R}_1{\bf R}_2} ({\bf q}) = \nonumber\\
&\frac{1}{N}\sum_{{\bf R}_1{\bf R}_2 {\bf R}_0} e^{i{\bf q}\cdot {\bf R}_0} e^{-i{\bf k}_1\cdot {\bf R}_1} e^{i{\bf k}_2\cdot {\bf R}_2}{\bf O}^{l_1l_2,\sigma}_{{\bf R}_1+{\bf R}_0,{\bf R}_2+{\bf R}_0} ({\bf q})\nonumber\\
&=\delta_{{\bf k}_1,{\bf k}_2-{\bf q}+{\bf G}_0} {\bf O}^{l_1l_2,\sigma}_{{\bf k}_1{\bf k}_2} ({\bf q}),
\end{align}
with 
\begin{align}
 {\bf O}^{l_1l_2,\sigma}_{{\bf k}_1{\bf k}_2} ({\bf q})&=\frac{1}{N}\sum_{{\bf R}_1{\bf R}_2}
 e^{-i{\bf k}_1\cdot {\bf R}_1} {\bf O}^{l_1l_2,\sigma}_{{\bf R}_1{\bf R}_2} ({\bf q})e^{i{\bf k}_2\cdot {\bf R}_2}
 \nonumber\\
 &=\frac{1}{\sqrt{V}}\int d{\bf r} \; e^{-i{\bf q}\cdot {\bf r}}\phi^*_{l_1{\bf k}_1}({\bf r})\phi_{l_2{\bf k}_2}({\bf r}) 
 .\label{OM}
\end{align}
We thus obtained the expression for the response function in an orbital basis that can be used for model calculations.

For a non-interacting system, the Bloch function basis set is usually prefered. It is useful to see how the oscillator matrix elements come in this representation. The time evolution of the operators in \eref{Resp0} in this case are trivial and one can evaluate the four-point correlation function to obtain the following expression for  the response function ($SU(2)$ symmetric case) in Matsubara-frequency space
\begin{align}
\chi^{0,ph}_{{\bf G}{\bf G}^{\prime}}&({\bf q},\nu_n)=-\frac{1}{N}\sum_{{\bf k}\sigma}\sum_{nm}
\frac{f(\epsilon_{m,{\bf k}-{\bf q}})-f(\epsilon_{n,{\bf k}})}
{\hbar\nu_n+\epsilon_{m,{\bf k}-{\bf q}}-\epsilon_{n,{\bf k}}}\nonumber\\ &\times 
\langle \psi_{n{\bf k}-{\bf q}}|e^{-i({\bf q}+{\bf G})\cdot \hat{{\bf r}}}|\psi_{m{\bf k}}\rangle
\langle \psi_{m{\bf k}}|e^{i({\bf q}+{\bf G}^{\prime})\cdot \hat{{\bf r}}}|\psi_{n{\bf k}-{\bf q}}\rangle,\label{Res}
\end{align}
where $\hat{\bf r}$ denotes the position operator, $\psi_{n{\bf k}}$ are the Bloch states, and $\chi^{0,ph}_{{\bf G}{\bf G}^{\prime}}({\bf q},\nu_n)\equiv \chi^{0,ph}({\bf q}+{\bf G},{\bf q}+{\bf G}^{\prime},\nu_n)$. Hereafter, we drop the reciprocal lattice vectors. They can easily be added when necessary. 

For a weakly to moderately correlated system it is useful to begin with \eref{Res} and use the perturbation expansion to consider correlation effects. In this approach, \eref{Res} should be rewritten in the orbital basis which is the natural basis for including interaction effects.
Expanding the Bloch functions in an atomic-like orbital basis set, namely, 
\begin{align}
\psi_{n{\bf k}}({\bf r}) =\sum_{l}  a^l_{n{\bf k}}\phi_{l{\bf k}}({\bf r})
\end{align}with the definition $a^l_{n{\bf k}}=\langle \phi_{l{\bf k}}|\psi_{n{\bf k}}\rangle$,
the matrix element for charge susceptibility can be rewritten as
\begin{align}
\langle \psi_{n{\bf k}-{\bf q}}|&e^{-i{\bf q}\cdot \hat{\bf r}}|\psi_{m{\bf k}}\rangle=
 \sum_{l_1l_2} a^{l_1*}_{n{\bf k}-{\bf q}}a^{l_2}_{m{\bf k}}{\bf O}_{{\bf k}}^{l_1l_2}({\bf q}),
\end{align}
where the oscillator matrix elements in the orbital basis are given by \eref{OM} with the short-hand (SU(2) symmetric case)
\begin{align}
{\bf O}_{{\bf k}}^{l_1l_2}({\bf q})={\bf O}_{{\bf k}-{\bf q},{\bf k}}^{l_1l_2,\sigma}({\bf q}).
\end{align}
Using this expression in the response function, one finds
\begin{align}
\chi^{0}_{ph}({\bf q},\nu_n)=\frac{1}{N}\sum_{{\bf k}\sigma}&\sum_{l_1l_2,l_3l_4}[{\bm \chi}^{0}_{ph}({\bf q},\nu_n)]_{{\bf k}}^{l_1l_2,l_3l_4}\nonumber\\ &\times
{\bf O}_{{\bf k}}^{l_1l_2}({\bf q}) {\bf O}_{{\bf k}}^{l_3l_4}(-{\bf q}),\label{ChargeSus}
\end{align}
where we defined the \emph{generalized susceptibility} by
\begin{align}
[{\bm \chi}^{0}_{ph}({\bf q},i\nu_n)]_{{\bf k}}^{l_1l_2,l_3l_4} \equiv &-\sum_{nm}
\frac{f(\epsilon_{m,{\bf k}-{\bf q}})-f(\epsilon_{n,{\bf k}})}
{i\nu_n+\epsilon_{m,{\bf k}-{\bf q}}-\epsilon_{n,{\bf k}}}\nonumber\\ &\times 
a^{l_1*}_{n{\bf k}-{\bf q}}a^{l_2}_{m{\bf k}}a^{l_3*}_{m{\bf k}}a^{l_4}_{n{\bf k}-{\bf q}}.\label{Chi0}
\end{align}
It describes the independent propagation of a p-h excitation and is  essential to study charge and magnetic  response functions in interacting systems (see below). It appears in the response to an external field and it solely depends on the electronic structure. This is the quantity that is usually computed and analyzed in the literature, instead of the response function \eref{ChargeSus}. In other words, the oscillator matrix elements are usually neglected. {\it Note that when orbital indices are not explicitly written, generalized susceptibilities are identified by a bold symbol to emphasize the tensor character. The same symbol but without boldface is used for the corresponding observable susceptiblities.}

The generalized susceptibility of  an interacting system is called the dressed generalized susceptibility, ${\bm \chi}_{ph}(Q)$. It is different from ${\bm \chi}^0_{ph}(Q)$ because 
propagating p-h pairs interact with their environment and with
each other. In a perturbation approach, the dressed generalized susceptibility  can be decomposed into  bare susceptibility, ${\bm \chi}^{0}_{ph}(Q)$, and vertex corrections, ${\bf X}_{ph}(Q)$, where we defined $Q \equiv ({\bf q},\nu_n)$  as the momentum-frequency four-vectors.  

 With the help of the irreducible vertex functions in the density channel, ${\bm \Gamma}^{irr,d}(Q)={\bm \Gamma}^{irr,\uparrow \downarrow}(Q)+{\bm \Gamma}^{irr,\uparrow \uparrow}(Q)$, the equation for the generalized  dressed susceptibility in this channel is~\cite{Bickers2004, PhysRevLett.117.137001}
\begin{align}
{\bm \chi}_{ph}^{d}&(Q) ={\bm \chi}^{0}_{ph}(Q)+{\bf X}_{ph}^{d}(Q)=\nonumber\\
&{\bm \chi}^{0}_{ph}(Q)-{\bm \chi}_{ph}^{d}(Q)
{\bm \Gamma}^{irr,d}(Q){\bm \chi}^{0}_{ph}(Q),
\label{eq:BSSus}
\end{align}
which can be rewritten as
\begin{align}
{\bm \chi}_{ph}^{d}(Q) ={\bm \chi}^{0}_{ph}(Q)[{\bm 1}+
{\bm \Gamma}^{irr,d}(Q){\bm \chi}^{0}_{ph}(Q))]^{-1}.
\label{eq:BSSus1}
\end{align}
Note that at the above two equations a summation over all internal wave vectors, frequencies and orbital indices is assumed. The Generalized susceptibilities are tensor quantities. 

 In order to obtain the observable charge response, the external legs of  the generalized susceptibility ${\bm \chi}_{ph}^{d}(Q)$ should be closed, just as the non-interacting case \eref{ChargeSus}, by the oscillator matrix elements, \eref{OM}. Therefore, the observable dressed correlation function is given by an equation similar to \eref{ChargeSus} where  ${\bm \chi}^{0}_{ph}({\bf q},\nu_n)$ is replaced by  ${\bm \chi}^{d}_{ph}({\bf q},\nu_n)$. 
Furthermore, as discussed in Appendix \ref{AppC}, the oscillator matrix elements also appear in the interaction matrix elements. While for local interactions one can absorb these matrix elements in the definition of the interaction, further-neighbor interactions necessitate extra phase factors that come from the oscillator matrix elements. 
 
 Inspecting the observable response function of the system in the normal phase provides the means to characterize a phase transition and identify the responsible driving mechanism. The response function becomes singular at a continuous (second order) phase transition into an ordered phase.  Since the oscillator matrix elements are temperature independent and non-singular, the singularity of the observable dressed correlation function  coincide with the singularity of the ${\bm \chi}^{d}_{ph}({\bf q},\nu_n)$. From \eref{eq:BSSus1} one can see that ${\bm \chi}^{d}_{ph}({\bf q},\nu_n)$ becomes singular once  $-{\protect \bm {\Gamma }}^{irr,   d}({\bf q}){\protect \bm {\chi }}_{ph}^0({\protect \bf q}, i\nu _n=0)$ has an eigenvalue equal to one. 
Hence, the  distance from a density transition is commonly measured by  the largest (dimensionless) eigenvalue of $-{\protect \bm {\Gamma }}^{irr,   d}({\bf q}){\protect \bm {\chi }}_{ph}^0({\protect \bf q}, i\nu _n=0)$. This is the so-called density Stoner factor, $\alpha_S^{d}({\bf q})$. An instability occurs once the Stoner factor becomes unity.  The corresponding wave-vector is the wave-vector of the instability. The Stoner factor analysis describes the tendency of the ground state towards a density or magnetic instability, independently of the external field. 

\subsection{Random phase approximation}

As one can see from \eref{ChargeSus}, the connection between theoretical and experimental results requires a calculation of the oscillator matrix elements.  The equation for the oscillator matrix elements,  \eref{OSReal}, can be rewritten as
\begin{align}
{\bf O}^{l_1l_2,\sigma}_{{\bf R}_1{\bf R}_2} &({\bf q})=\exp[{-i{\bf q}\cdot ({\bf R}_1+{\bf r}_{\alpha_1}+{\bm \delta}/2)}]\nonumber\\\times\frac{1}{\sqrt{V}}&\int d{\bf r} 
e^{-i{\bf q}\cdot {\bf r}} 
\phi^*_{ l_1,\sigma}({\bf r}+{\bm \delta}/2) \phi_{ l_2,\sigma}({\bf r}-{\bm \delta}/2),\label{OSReal2}
\end{align}
where ${\bm \delta}\equiv ({\bf R}_2+{\bf r}_{\alpha_2})-({\bf R}_1+{\bf r}_{\alpha_1})$. Note that the second line at the above equation \emph{only} depends on the spatial distance of the orbital centers and is translationally invariant. It is clear that one cannot get rid of the phase in the prefactor by changing the gauge since that phase depends on the center of mass coordinate. 

In practice, a commonly used approximation is the so-called random phase approximation (RPA). 
In this approximation, one assumes a static irreducible vertex, 
and only oscillator matrix elements with identical orbital centers and character are kept, i.e., 
\begin{equation}
 {\bf O}^{l_1l_2,\sigma}_{{\bf R}_1{\bf R}_2} ({\bf q}) \propto  \exp[-i{\bf q}\cdot ({\bf R}_1+{\bf r}_{\alpha})] \delta_{l_1l_2} \delta_{{\bf R}_1,{\bf R}_2},\label{RPAOS0}
\end{equation}
where orbital $l_1$ belongs to atom $\alpha$.  
This approximation is only justified in the small wave-vector ${\bf q} \rightarrow {\bm 0}$ or very large wave-length limit where the inner structure of the unit cell is irrelevant. Actually, the oscillator matrix elements with nearest-neighbor $l_1$ and $ l_2$ are not negligible for finite ${\bf q}$. Indeed, the hopping matrix elements are given by a similar overlap integral where the phase factor is replaced by the Laplacian, see \eref{Ham0}, which oscillates much faster. 

Within this approximation, the oscillator matrix element in Fourier space Eq.~\eqref{OM} is proportional to
\begin{equation}
 {\bf O}^{l_1l_2,\sigma}_{{\bf k}} ({\bf q}) \propto  \exp(-i{\bf q}\cdot {\bf r}_{\alpha}) \delta_{l_1l_2} ,\label{RPAOS}
\end{equation}
which allows to perform  the ${\bf k}$-summation in \eref{ChargeSus}. Hence, the resulting observable response function  is given by
\begin{align}
\chi^{0}_{ph}({\bf q},\nu_n)=&\sum_{l_1l_2}[{\bm \chi}^{0,RPA}_{ph}({\bf q},\nu_n)]_{l_1l_1,l_2l_2}\nonumber\\ &\times
\exp[-i{\bf q}\cdot ({\bf r}_{\alpha}-{\bf r}_{\beta})],\label{RPAChargeSus}
\end{align}
where $l_1$ and $l_2$ orbitals belong to atoms $\alpha$ and $\beta$ respectively.  In many studies the phase factor in \eref{RPAChargeSus} is incorrectly neglected. The so-called irreducible
RPA generalized  susceptibility in the above equation can be rewritten in terms of the Green's function as
\begin{align}
[{\bm \chi}^{0,RPA}_{ph}(Q)]_{l_1l_1;l_2l_2} 
=- (\frac{k_BT}{N})\sum_K {\bf G}_{K-Q,l_1l_2}{\bf G}_{K,l_2l_1},\label{eq:bareSusRPA}
\end{align}
where $K \equiv ({\bf k},\omega_m)$.
It is worth emphasizing that \eref{RPAOS0}, \eref{RPAOS} and  \eref{RPAChargeSus} are gauge independent. 
The difference between systems with symmorphic space-group symmetry and those with non-symmorphic space-group symmetry appears here. For systems with symmorphic space group, the generalized susceptibility has the same symmetries as the response function, while for systems with non-symmorphic space group it is necessary to take into account the oscillator matrix elements to obtain the observed symmetry of the response function.

The correct observable symmetries are also recovered in the RPA approximation for the dressed observable susceptibility. Closing the external legs of the dressed generalized susceptibility by \eref{RPAOS} yields an equation similar to \eref{RPAChargeSus} where  ${\bm \chi}^{0}_{ph}({\bf q},\nu_n)$ is replaced by  ${\bm \chi}^{d}_{ph}({\bf q},\nu_n)$. The lowest correction to the bare observable correlation function is 
\begin{align}
-& \sum_{l_1\ldots l_6} e^{-i{\bf q}\cdot {\bf r}_{\alpha}}[{\bm \chi}^{0,RPA}_{ph}({\bf q},\nu_n)]_{l_1l_1,l_3l_4}\nonumber\\ &\times
[{\bm \Gamma}^{d,irr}_{RPA}({\bf q},\nu_n)]_{l_3l_4,l_5l_6}[{\bm \chi}^{0,RPA}_{ph}({\bf q},\nu_n)]_{l_5l_6,l_2l_2}
e^{i{\bf q}\cdot {\bf r}_{\beta}},\label{RPADressedChargeSus}
\end{align}
where $l_1$ and $l_2$ orbitals belong to atoms $\alpha$ and $\beta$ respectively and ${\bm \Gamma}^{d,irr}_{RPA}$ denotes the irreducible vertex function in the RPA approximation. In the case of a local interaction, $l_3 \ldots l_6$ orbitals belong to same atom $\gamma$. By multiplying the above equation with $\exp[-i{\bf q}\cdot ({\bf r}_{\gamma}-{\bf r}_{\gamma})]$ one can show that resulting expression has same symmetry as the observable bare susceptibility and therefore the interacting correlation function has it as well. Extension to the case of non-local interaction is straightforward.  As discussed in Appendix \ref{AppC},  non-local interaction matrix elements acquire extra phase factors that come from the oscillator matrix elements. 

\section{Nonsymmorphic symmetry operations and Hamiltonian transformation}\label{Transformation}
In this section we discuss how the basis dependent ${\bf H}_{{\bf k}}$, \eref{Ham1}, transforms under point group symmetry operations of the lattice. This is a necessary step to demonstrate how oscillator matrix elements allow response functions to exhibit, at certain wave vectors, symmetries that are absent from ${\bf H}_{{\bf k}}$. We restrict ourselves to the interesting case of the iron-pnictide superconductors with a non-symmorphic point group. With $16$ Fe-$d$ and As-$p$ orbitals in the unit cell, the Hamiltonian in the orbital basis is given by ${\bf {\mathcal H}}=\sum_{{\bf k}} {\bm \phi}^{\dagger}_{{\bf k}}{\bf H}_{\bf k} {\bm \phi}_{{\bf k}}$,
where ${\bm \phi}_{\bf k} \equiv (c_{1,{\bf k}},\ldots, c_{16,{\bf k}})^T$.

\begin{figure}
\centering{
	\includegraphics[width=0.75\linewidth,clip=]{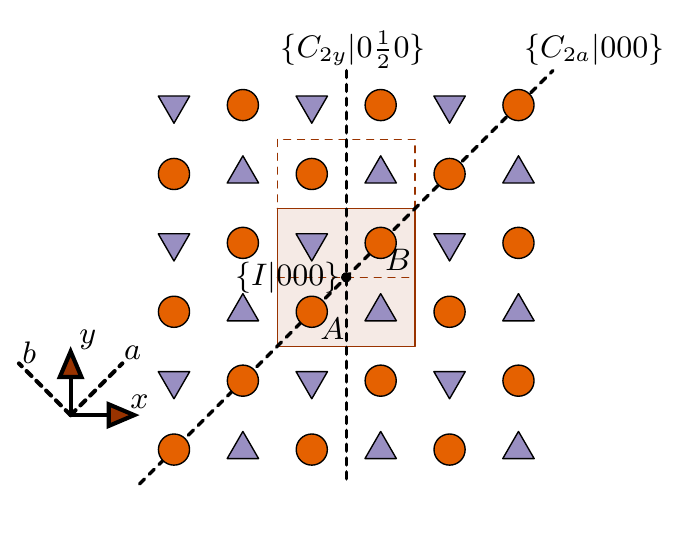}}
	\caption{(Color online) Sketch of the lattice structure of a trilayer Fe-As unit. The iron and the pnictide atoms are shown respectively by circles and triangles. Down triangles indicate that the As is below the plane and up triangles that As is above the plane. The inversion center is located in the middle of an Fe-Fe link. The three generators of the $P4/nmm$ space group, $\{C_{2y}|0\frac{1}{2}0 \}$, $\{C_{2a}|000 \}$ and inversion $\{I|000\}$, are illustrated. The dashed square shows how the unit cell in the shaded area transforms under $\{C_{2y}|0\frac{1}{2}0 \}$. 
	}\label{fig1}
\end{figure}

 In general, a non-translation member of the space group can be described by $\{g|{\bm \tau}\}$, where $g$ is a point group operation (rotation, reflection, etc.), which keeps at least one point (the coordinate origin) invariant, followed by a \emph{half-integer} translation by vector $\bm \tau$.~\cite{PhysRevB.88.134510} The half-integer translation vectors depend on the choice of the origin and axis.  They cannot be eliminated by shifting the origin for screw axis and glide-plane operations. A space group including these symmetries  is called non-symmorphic.  
\fref{fig1} sketches the lattice structure of a trilayer Fe-As unit and three generators of the $P4/nmm$ group (see appendix \ref{AppA} for a full table of the symmetry operations). The origin is chosen to be on  the inversion center in the middle of an Fe-Fe link.  We consider the three generators $\{C_{2y}|0\frac{1}{2}0 \}$, $\{C_{2a}|000 \}$ and inversion $\{I|000\}$, where $C_{n{\bf r}}$ is an anti-clockwise rotation through $2\pi/n$ radians about the axis labeled by ${\bf r}$. The $\{C_{2y}|0\frac{1}{2}0 \}$ operation is a screw axis operation where a rotation is followed by a translation along the rotation axis. For a non-symmorphic space group, the non-translation elements $\{g|{\bm \tau}\}$ do not form a group, since their combination may turn out to be a pure lattice translation.  However,  the symmorphic elements $\{g|{\bm 0}\}$ by themselves form a group. 

Consider an atom $\alpha$ at the $i$th unit cell located at ${\bf R}_i+{\bf r}_{\alpha}$.
% where ${\bf r}_{\alpha}$ denotes the position of the atom within unit cell. 
 Under a symmetry operation, $\{g|{\bm \tau}\}$, this atom transfers to $g{\bf R}_i+g{\bf r}_{\alpha}+{\bm \tau}$. This is accompanied by a rotation in the associated orbital space along with a possible atom index exchange $\alpha \rightarrow \alpha^{\prime}$ if there is more than one atom of the same kind in the unit cell. The peculiarity of a non-symmorphic operation is that the new position may be in a unit cell adjacent to  $g{\bf R}_i$. Defining ${\bf L_\alpha}$ as a real space lattice vector which connects atoms $\alpha^{\prime}$ in these two unit cells, 
we have $g{\bf r}_{\alpha}+{\bm \tau} = {\bf r}_{\alpha^{\prime}}+{\bf L}_\alpha$. The vector ${\bf L}_{\alpha}$ depends on the position of atom $\alpha$ and on the vector $\bm \tau$. For example, under $\{C_{2y}|0\frac{1}{2}0\}$ the $B$-site of the unit cell at origin transfers to $A$-site of the neighboring unit cell in the $y$-direction while the $A$-site transfers to  $B$-site of the  same unit cell (see \fref{fig1}). Thus, for this symmetry operation  ${\bf L}_\alpha={\bm 0}$ for sublattice $A$ and ${\bf L}_\alpha={\bf a}_2$ for sublattice $B$, where ${\bf a}_j$ denotes the real-space lattice unit vector in $j$th direction. In Fourier space, these vectors ${\bf L}_\alpha$ introduce atom-dependent extra phases.

As shown in  appendix \ref{AppB}, under a symmetry operation, $\{g|{\bm \tau}\}$, the electron annihilation  operators in the orbital basis $\phi_{{\bf k}}$  transform as 
$\{g|{\bm \tau}\}\phi_{\bf k}={\bf U}^{\prime}_{g{\bf k},{\bf L}}{\bf U}_g\phi_{g{\bf k}}$
where   ${\bf U}_g$ is a $16\times 16$ unitary matrix that describes the effect of the $\{g|{\bm \tau}\}$ operation in the orbital basis and where the action of $g$ on ${\bf k}$ is represented by a $3\times 3$ orthogonal matrix. The existence of the above mentioned ${\bf L}_{\alpha}$ vectors leads to a diagonal $16\times 16$ unitary matrix  ${\bf U}^{\prime}_{g{\bf k},{\bf L}}$  with element 
$\exp(-ig{\bf k}\cdot {\bf L}_{\alpha})$. 
Therefore, the invariance of the Hamiltonian takes the form 
\begin{equation}
{\bf H}_{g{\bf k}}={\bf U}^{\dagger}_g{\bf U}^{\prime \dagger}_{g{\bf k},{\bf L}}{\bf H}_{{\bf k}}
{\bf U}^{\prime}_{g{\bf k},{\bf L}}{\bf U}_g.\label{Tran1}
\end{equation}
which is different from the result for a symmorphic operation, i. e.,
 \begin{equation}
 {\bf H}_{g{\bf k}}={\bf U}^{\dagger}_g{\bf H}_{{\bf k}}{\bf U}_g. \label{Tran2}
 \end{equation}

For iron-based superconductors,  the  usual relation, \eref{Tran2}, is recovered under the $\{I|000\}$  and $\{C_{2a}|000 \}$ generators where the extra phase factors are absent.  However under $\{C_{2y}|0\frac{1}{2}0 \}$ only one sublattice obtains the extra phase, $\exp[-i2\pi (g{\bf k})_y]$, where $(g{\bf k})_y$ is the $y$-component of the wave-vector in reciprocal lattice vector units. 
This implies that for this operation \eref{Tran2} is valid only for the intra-sublattice part of the Hamiltonian, where the extra phases are canceled out, and not for the whole matrix. 
The same is true for all the symmetry operations of the group with non-zero $\bm \tau$. 
It is not possible to eliminate these phases with a gauge transformation for the same reason that it is not possible to eliminate the half-integer translation with a shift of origin. These symmetry properties are clarified further using a minimal model Hamiltonian in appendices \ref{TBHamiltonian} and \ref{Sec:SymmInPlane}.

Given the above symmetry properties for ${\bf H}_{\bf k}$, a key question is how the observables recover the full symmetry of the lattice in the absence of any symmetry breaking phase transition. Band energies and their degeneracy are clearly invariant under all symmetry operations. This can be seen from the invariance of ${\rm det}(\omega{\bm 1}-{\bf H}_{\bf k})$ under symmetry operations and the fact that the determinant of the matrix product is equal to the product of the determinants.  
Another important quantity, connected to electron charge density, is the electron momentum distribution function, which is given by $n({\bf k})=1+2{\rm Re}\sum_{\omega_m} {\rm Tr}[{\bf O}_{{\bf k}}({\bf q}={\bm 0}){\bf G}({\bf k},i\omega_m)]$. Here, ${\bf G}({\bf k},i\omega_m)=[(i\omega_m+\mu) {\bm 1}-{\bf H}_{\bf k}]^{-1}$ denotes the electron propagator. Since ${\bf O}^{l_1l_2}_{{\bf k}}({\bf q}={\bm 0}) \propto \delta_{l_1l_2}$ as can be seen from \eref{OM},  only intra-orbital components contribute in the trace, hence this quantity is also invariant, in usual sense, under all symmetry operations. The same is true for diagonal components of the spectral function. 

Coming back to the suseptibility, both the generalized susceptibility and the observable ones behave the same way under a symmorphic point-group operation. Such a transformation permutes atomic orbitals into one another and changes positions following $\textbf{r}_\alpha \rightarrow g\textbf{r}_\alpha$. In Fourier components this translates into $\textbf{q} \rightarrow g^{-1}\textbf{q}$ and the phase factor coming from the oscilator matrix elements in the calculation of the observable susceptibility Eq.(\ref{RPAChargeSus}) has the same property. 

For a non-symmorphic operation on the generalized susceptibility, there is an additional phase $\exp[-i2\pi (g{\bf k})_y]$, where $(g{\bf k})_y$ is the $y$-component of the wave-vector in reciprocal lattice vector units, that depends on the position of the atom in the unit cell. The phase factor in the transformation from the generalized susceptibility to the observable susceptibility \eref{RPAChargeSus} has an analogous dependence on position within the unit cell that compensates that phase and $C_{4z}$ symmetry at $q_z=0$ is recovered.

In the following section we illustrate further the more subtle case of the symmetry properties of the generalized susceptibilities by performing explicit calculations for LiFeAs. We show that the generalized susceptibilities do not have $C_{4z}$ symmetry at $q_z=0$.  The full symmetries are recovered only when the generalized susceptibilities are embeded back in the continuum using appropriate matrix elements between basis states, as discussed above. In appendices \ref{TBHamiltonian} and \ref{Sec:SymmInPlane} we perform the full calculation for a minimal tight-binding Hamiltonian to illustrate in detail what happens.

\section{Generalized susceptibility of $\rm{\bf LiFeAs}$}\label{Result}
Here, we present some results for the in-plane generalized susceptibilities in the RPA approximation. Although our discussion is general, to be specific we present results on LiFeAs.~\cite{PhysRevB.78.060505}
We are particularly interested in four-fold symmetry in the normal state. We show explicitly that non-symmorphic operations lead to generalized susceptibilities that do not have $C_{4z}$ symmetry even at $q_z=0$. This symmetry is recovered in observable susceptibilities if we take into account oscillator matrix elements.

We work in the two-Fe unit cell and all Fe-$3d$ and As-$4p$ orbitals are considered in the calculations. 
In what follows, we focus on the $d$ orbitals  of Fe-$1$ (Fe-$2$) (on $A$ and $B$ sublattices respectively) that have large  weight around the Fermi level: $d_{x^2-y^2}$ will be referred as $2 (7)$ and $d_{xz}$ and $d_{yz}$ orbitals as $4 (9)$ and $5 (10)$. Note that in the one-Fe unit cell with a coordinate system rotated by $\pi/4$ around the $z$-axis relative to the two-Fe coordinate system, the $d_{x^2-y^2}$ orbital becomes $d_{xy}$.

\begin{figure}
\centering{
		\begin{tabular}{ccc}
			\includegraphics[width=0.33\linewidth,clip=]{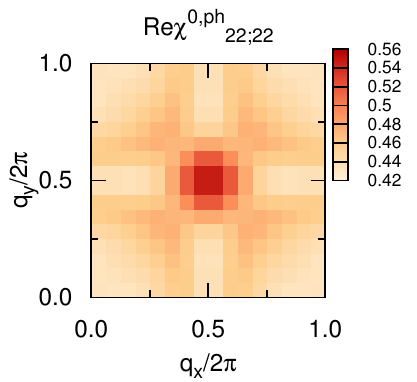} &
			\includegraphics[width=0.33\linewidth,clip=]{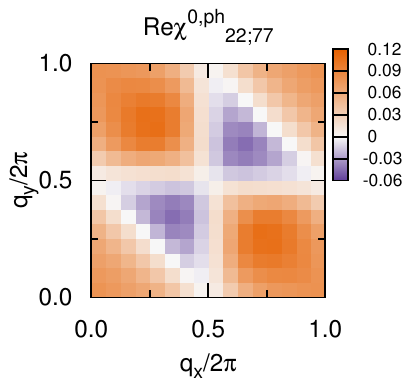} &
			\includegraphics[width=0.33\linewidth,clip=]{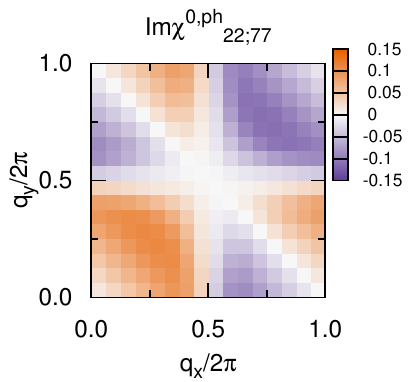}			
		\end{tabular}}
		\caption{(Color online) Real and imaginary parts of the $d_{x^2-y^2}$ intra-orbital, intra-sublattice $22;22$ (left) and inter-sublattice $22;77$ (middle and right) component of the in-plane bare generalized susceptibility at the lowest bosonic frequency, $[{\bm \chi}^{0,RPA}_{ph}({\bf q}, \nu_n=0)]$, of LiFeAs at $k_B T = 0.01$ eV in the p-h channel.  The Kohn-Sham eigenstates are projected to atomic-like Wannier basis to construct the Hamiltonian in the orbital basis.~\cite{PhysRevB.80.085101} For $K$ summation in \eref{Chi0} we have used a $32\times 32\times 16$ $k$-mesh and $1024$ positive frequencies. }\label{fig2}
\end{figure}
The left panel of \fref{fig2} shows  the in-plane bare generalized  susceptibility $[{\bm \chi}^{0,RPA}_{ph}]_{22;22}$ ($=[{\bm \chi}^{0,RPA}_{ph}]_{77;77}$) of LiFeAs  at the lowest bosonic Matsubara frequency. It is the dominant component of the bare susceptibility with incommensurate peaks around the $M$ point, which comes from nesting between the hole and electron
pockets. (We use the term incommensurate in a general sense to distinguish from high symmetry points such as the M point. With this definition, a wave vector $(\pi/3,\pi)$ is incommensurate.) The $d_{xz(yz)}$ intra-orbital components, $[{\bm \chi}^{0,RPA}_{ph}]_{44;44}$ and $[{\bm \chi}^{0,RPA}_{ph}]_{55;55}$, show commensurate peaks at the $M$ point (not shown).~\cite{PhysRevLett.117.137001, PhysRevB.93.241116} These  components are obtained from the convolution of two intra-sublattice propagators and therefore they are invariant under all symmetry operations. The $d_{x^2-y^2}$ component has four-fold symmetry while the $d_{xz(yz)}$ components are degenerate and are transformed into each other under four-fold operations. 

However,  a general component of the in-plane bare generalized susceptibility does not have four-fold symmetry because the corresponding symmetry operator is non-symmorphic.
For example, the middle and right panels of \fref{fig2} shows the real and imaginary parts of  the in-plane $[{\bm \chi}^{0,RPA}_{ph}]_{22;77}$ component as an example of components that include inter-sublattice propagators. 
Since ${\bf G}_{27}(K+Q)$ and ${\bf G}_{72}(K)$ in the susceptibility  involve different wave vectors and Fe sites, the non-symmorphic phases do not cancel. These components have a lower symmetry than those including only intra-sublattice propagators.  This leads to a generalized susceptibility without four-fold symmetry. 

\begin{figure}
\centering{
\begin{tabular}{cc}
\includegraphics[width=0.49\linewidth,clip=]{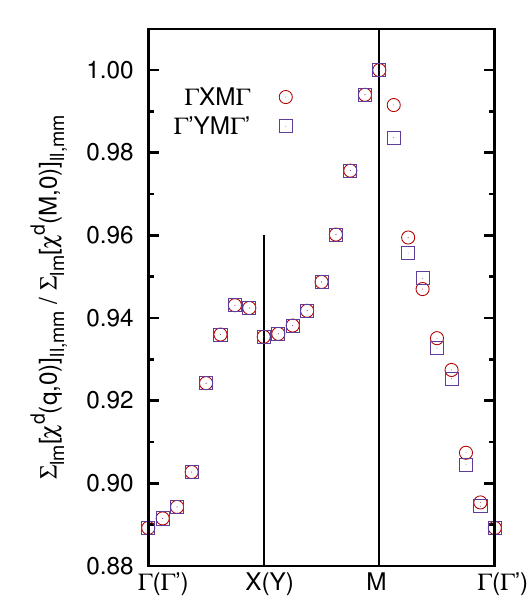} &
\includegraphics[width=0.49\linewidth,clip=]{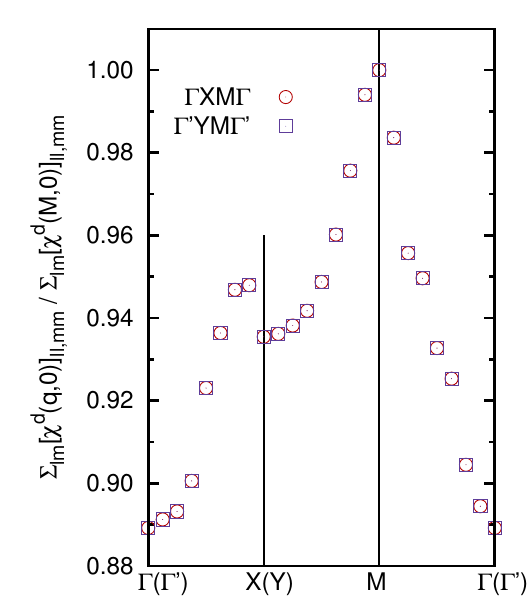}
\end{tabular}}
\caption{(Color online) RPA observable dressed susceptibility in density channel of LiFeAs  at $k_BT=  0.01$~eV,  without (left) and with (right) the oscillator matrix elements (see \eref{RPAChargeSus}).  The fourfold symmetry is absent for wave vectors within the BZ (compare $\Gamma-M$ and $\Gamma^{\prime}-M$ paths on the left panel) if one neglects the phases due to the oscillator matrix elements (left panel). This symmetry is recovered by including these phases (right panel).  
The $\Gamma-M$ path is along the primary diagonal (positive slope) while  $\Gamma^{\prime}-M$ path is along the secondary diagonal (negative slope). The screened interaction parameters used in the RPA dressing are $J_s=0.1U_s$ and $U_s=1.32$~eV.~\cite{PhysRevLett.117.137001} The inter-orbital interaction and pair hopping are determined assuming spin-rotational symmetry.  }\label{fig2-1}
\end{figure}

\fref{fig2-1} demonstrates $\sum_{l,m}[{\protect \bm {\chi }}_{ph}^0({\protect \bf q})]_{ll,mm}$ normalized by $\sum_{l,m}[{\protect \bm {\chi }}_{ph}^0(M)]_{ll,mm}$ along a high symmetry path (left panel). Despite the fact that the bare charge susceptibility along the main axes shows $C_4$ symmetry
the susceptibility along the primary $\Gamma-M$ and secondary  $\Gamma'-M$  diagonals are slightly different. However, taking into account the role played by oscillator matrix elements through \eref{RPAChargeSus}, leads to a response function which has four-fold symmetry (right panel).

Finally, in spin-fluctuation mediated pairing, the density/magnetic fluctuations contribute to the pairing interaction. These fluctuations are expressed in terms of the dressed generalized susceptibilities and vertex functions in density and magnetic channels.~\cite{PhysRevLett.117.137001, PhysRevB.93.241116} Hence, the anisotropy of the generalized susceptibility would influence the superconducting phase as well. Solving the Eliashberg equation yields the leading gap functions which either preserves the full symmetry or has a lower symmetry.  Imposing an artificial four-fold symmetry on the generalized susceptibilities, as is done in one-iron unit cell studies, could impact the competition between the leading gap functions and could switch their order.

\section{Conclusion}\label{Con}
In conclusion, we clarified the effects of the non-symmorphic symmetry operations on the Hamiltonian and on the response and correlation functions in the orbital basis. These symmetry operations include half-integer translations that introduce extra phases when applied to the Hamiltonian. As an example, the four-fold rotational symmetry of the iron-based pnictide LiFeAs that is expected for $q_z=0$ does not exist for the non-diagonal part of the Hamiltonian.  We showed that while single-particle eigenvalues maintain full four-fold symmetry, that symmetry is only recovered for observable response functions when one takes into account the symmetry properties of both the generalized susceptibility and of the oscillator matrix elements. This was further clarified using a minimal tight-binding Hamiltonian for iron pnictides, which have a non-symmorphic space group. 

\begin{acknowledgments}
	R.~N is grateful to G.~ Kotliar for discussions which led to this work and to J. Gukelberger, H. Kontani, O. Vafek and I. Mazin for discussions. We are also thankful of A.~Linscheid and P.~J.~Hirschfeld for a careful reading of the manuscript and for their useful comments. This work was supported by the Natural Sciences and Engineering Research Council of Canada (NSERC) under grant RGPIN-2014-04584, and by the Research Chair in the Theory of Quantum Materials. Simulations were performed on computers provided by the Canadian Foundation for Innovation, the Minist\`ere de l'\'Education des Loisirs et du Sport (Qu\'ebec), Calcul Qu\'ebec, and Compute Canada.
\end{acknowledgments}

\appendix
\section{Interaction matrix elements}\label{AppC}
The interaction term is given by 
\begin{align}
H_{int} &= \frac{1}{2}\sum_{\sigma \sigma^{\prime}} \nonumber \\ \frac{1}{V}&\int d{\bf r} d{\bf r}^{\prime}
\Psi^{\dagger}_{\sigma}({\bf r}) \Psi^{\dagger}_{\sigma^{\prime}}({\bf r}^{\prime})\mathcal{V}({\bf r}-{\bf r}^{\prime})
\Psi_{\sigma^{\prime}}({\bf r}^{\prime})\Psi_{\sigma}({\bf r}).
\end{align}
Using the field operator expansion in the orbital basis, the interaction term can be rewritten as
\begin{align}
H_{int} = \frac{1}{2}&\sum_{\sigma \sigma^{\prime}} 
\sum_{ l_1 \ldots  l_4}\sum_{{\bf R}_1 \ldots {\bf R}_4}
\mathcal{V}_{{\bf R}_1{\bf R}_2;{\bf R}_3{\bf R}_4}^{l_1,l_2;l_3,l_4}\nonumber\\
&\hat{c}^{\dagger}_{{\bf R}_1 l_1,\sigma} \hat{c}^{\dagger}_{{\bf R}_2 l_2,\sigma^{\prime}}\hat{c}_{{\bf R}_3 l_3,\sigma^{\prime}}\hat{c}_{{\bf R}_4 l_4,\sigma},
\end{align}
with 
\begin{align}
\mathcal{V}_{{\bf R}_1{\bf R}_2;{\bf R}_3{\bf R}_4}^{l_1,l_2;l_3,l_4}\equiv \frac{1}{V}
\int &d{\bf r} d{\bf r}^{\prime} \phi^*_{{\bf R}_1 l_1}({\bf r})\phi^*_{{\bf R}_2 l_2}({\bf r}^{\prime})\nonumber\\
&\mathcal{V}({\bf r}-{\bf r}^{\prime})
\phi_{{\bf R}_3 l_3}({\bf r}^{\prime})\phi_{{\bf R}_4 l_4}({\bf r}).\label{V}
\end{align}

One can easily see that $\mathcal{V}_{{\bf R}_1{\bf R}_2;{\bf R}_3{\bf R}_4}^{l_1,l_2;l_3,l_4} = \mathcal{V}_{{\bf R}_1+{\bf R}_0,{\bf R}_2+{\bf R}_0;{\bf R}_3+{\bf R}_0,{\bf R}_4+{\bf R}_0}^{l_1,l_2;l_3,l_4}$. 
 What is normally used as an input parameters in model Hamiltonian studies or in the  LDA+DMFT calculation or an RPA analysis of the instability is the left-hand side of \eref{V}. It is clear that this interaction parameter depends on the choice of orbital (basis) and therefore its symmetry properties also depend on the basis. However, one can dress susceptibilities in the orbital basis, hence oscillator matrix elements only appear when closing the external legs of the dressed generalized susceptibilties to obtain observable susceptibilities.

\section{$P4/nmm$ space group}\label{AppA}
$P4/nmm$ is a non-symmorphic group. It has eight point group operations (see  Fig.~1 of main text). In addition to the eight point group operations, it also has an inversion, $\{I|000\}$. Together they form the  $16$ elements of $P4/nmm$. Note that this set is not closed under multiplication,  and  therefore,  as  is,  it  cannot  form  a  group. However, if one  defines  the  product  modulo  integer translations,  then  these  $16$  operations  form  a  group.

The first eight  symmetry operations can be generated by $\{C_{2y}|0\frac{1}{2}0\}$ and $\{C_{2a}|000\}$ as:
\begin{align*}
\{C_{2y}|0\frac{1}{2}0\}\{C_{2a}|000\}&=\{C^+_{4z}|0\frac{1}{2}0\},\\
\{C_{2a}|000\}\{C_{2y}|0\frac{1}{2}0\}&=\{C^-_{4z}|\frac{1}{2}00\},\\
\{C^+_{4z}|0\frac{1}{2}0\}\{C^+_{4z}|0\frac{1}{2}0\}&=\{C_{2z}|\frac{1}{2}\frac{1}{2}0\}+\{E|-1,0,0\},\\
\{C_{2y}|0\frac{1}{2}0\}\{C_{2z}|\frac{1}{2}\frac{1}{2}0\}&=\{C_{2x}|\frac{1}{2}00\}+\{E|-1,1,0\},\\
\{C_{2a}|000\}\{C_{2z}|\frac{1}{2}\frac{1}{2}0\}&=\{C_{2b}|\frac{1}{2}\frac{1}{2}0\},
\end{align*}
where the translational part $\{E|-1,0,0\}$ will be dropped hereafter. Note that the $b$-axis is in the Fe plane, so $C_{2b}$ transfers the As ions from above to below and vice versa, hence a half-integer translation is required to bring the lattice into itself. The other eight operations can be obtained by applying inversion  $\{I|000\}$ on the previous operators. 

\section{Effect of non-symmorphic operations}\label{AppB}
Let the basis set be $\phi_{l_{\alpha}}(  \mathbf{R}_{j}+\mathbf{r}_{\alpha})  $. Here, $\mathbf{R}_{j}$ denotes the position of the $j^{th}$
unit cell and $\mathbf{r}_{\alpha}$ the position of the atom $\alpha$ within the unit cell. The index $\alpha$ can take
four values, two of which are associated with Fe and two with As.
Associated with each atom, there is a set of atomic orbitals denoted by
$l_{\alpha}$ where for a given $\alpha$, $l_{\alpha}$ can take several values.
For example, if atom $\alpha$ is an Fe atom, then
$l_{\alpha}$ can take five values corresponding to $d$ orbitals while if
it is an As atom, then $l_{\alpha}$ can take
three values corresponding to the three $p$ orbitals.  In the main
text, the label $l$ for the \textquotedblleft orbital\textquotedblright%
\ labels both the atomic position and the orbital on the atom. Here we
temporarily make the distinction between the atomic orbital and the position
of the atom. 

The action of $\left\{  g|{\bm \tau}\right\}  $ is given by%
\begin{equation}
\left\{  g|{\bm \tau}\right\}  \phi_{l_{\alpha}}(  \mathbf{R}_{j}%
+\mathbf{r}_{\alpha})  =\phi_{l^{\prime}_{\alpha^{\prime}}}(  g\mathbf{R}_{j}%
+g\mathbf{r}_{\alpha}+{\bm \tau})
\end{equation}
where $(\alpha,l_{\alpha})$ transfers to $(\alpha^{\prime},l^{\prime}_{\alpha^{\prime}})$ under the symmetry operation. The operations act not only on spatial coordinates, they also act
on the atomic orbital itself. Take for example $C_{2y}$ for $g$, then it
would transform a $p_{z}$ orbital of As into $-p_{z}$ and would leave a Fe
$d_{xz}$ orbital unchanged. We also have to remember that the action of the
operation $g$ (that leaves the origin invariant) on the vectors $\mathbf{R}%
_{j}$ and $\mathbf{r}_{\alpha}$ is represented by the same $3\times3$ matrix. 
We
also have the result%
\begin{equation}
g\mathbf{r}_{\alpha}+{\bm \tau}=\mathbf{r}_{\alpha^{\prime}}+{\bf L}_{\alpha}%
\end{equation}
where $\mathbf{r}_{\alpha^{\prime}}$ is in the $g\mathbf{R}_{j}$ unit cell and
${\bf L}_{\alpha}$ is a real space lattice vector necessary to reach,
from $\mathbf{r}_{\alpha^{\prime}}$, the position $g\mathbf{r}_{\alpha}+{\bm \tau}$ that may
be in a unit cell adjacent to $g\mathbf{R}_{j}$. It depends on the
original position $\mathbf{r}_{\alpha}$. The position $\mathbf{r}
_{\alpha^{\prime}}$ will correspond to the same kind of atom, Fe or As, as the
original one, but may be displaced in the unit cell. A Fe atom originally on
sublattice $A$ may have moved to sublattice $B$ for example, in other words
$\alpha$ and $\alpha^{\prime}$ may be different, even if they both refer to an Fe atom.

The action of $\{g|{\bm \tau}\}$ on the orbitals $\phi_{l_{\alpha}}$ may be represented by a
unitary matrix. Hence, we may write%
\begin{equation}
\left\{  g|{\bm \tau}\right\}  \phi_{l_{\alpha}}(  \mathbf{R}_{j}%
+\mathbf{r}_{\alpha})  =U_{g}^{l_{\alpha};l^{\prime}_{\alpha^{\prime}}}%
\phi_{l_{\alpha^{\prime}}^{\prime}}(  g\mathbf{R}_{j}+\mathbf{r}%
_{\alpha^{\prime}}+{\bf L}_{\alpha})  \;
\end{equation}
where here $\alpha^{\prime}$ is given and a summation over the different values
$l^{\prime}_{\alpha^{\prime}}$ associated with atom $\alpha^{\prime}$ is implied.

We now move to Fourier transforms. We use the standard gauge because the Bloch Hamiltonian is periodic in this gauge. We comment on the other gauge whenever it is appropriate. We define
\begin{equation}
\phi_{\alpha l_{\alpha}}( \mathbf{k})  =\frac{1}{\sqrt{N}}\sum
_{{\bf R}_j}e^{i\mathbf{k\cdot R}_{j}}\phi_{l_{\alpha}}(  \mathbf{R}%
_{j}+\mathbf{r}_{\alpha}) .\label{STgauge}
\end{equation}
The action of $\left\{  g|{\bm \tau}\right\}  $ on $\phi_{\alpha l_{\alpha}}(
\mathbf{k})  $ follows if we also remember that under $g$ (seen simply
as a coordinate change for example), the scalar product must be invariant so
that $g\mathbf{k\cdot}g\mathbf{R}_{j}=\mathbf{k\cdot R}_{j}.$ We find,
\begin{widetext}
\begin{align}
\left\{  g|{\bm \tau}\right\}  \phi_{\alpha l_{\alpha}}(  \mathbf{k})  &=\frac{1}{\sqrt{N}}\sum_{{\bf R}_j}
e^{i\mathbf{k}\cdot\mathbf{R}_{j}}U_{g}^{l_{\alpha};l^{\prime}_{\alpha^{\prime}}}%
\phi_{l_{\alpha^{\prime}}^{\prime}}(  g\mathbf{R}_{j}+g\mathbf{r}_{\alpha}+{\bm \tau})  
=
\frac{1}{\sqrt{N}}
\sum_{{\bf R}_j}e^{ig\mathbf{k}\cdot g\mathbf{R}_{j}}U_{g}^{l_{\alpha};l^{\prime}_{\alpha^{\prime}}}%
\phi_{l_{\alpha^{\prime}}^{\prime}}(  g\mathbf{R}_{j}+\mathbf{r}_{\alpha^{\prime}}+\mathbf{L}_{\alpha})  \nonumber\\
&  =
\frac{1}{\sqrt{N}}
e^{-ig\mathbf{k}\cdot {\bf L}_{\alpha}}\sum_{{\bf R}_j}e^{ig\mathbf{k}\cdot(g\mathbf{R}_{j}+{\bf L}_{\alpha}%
)}U_{g}^{l_{\alpha};l^{\prime}_{\alpha^{\prime}}}%
\phi_{l_{\alpha^{\prime}}^{\prime}}(  g\mathbf{R}_{j}+\mathbf{r}_{\alpha^{\prime}}+{\bf L}_{\alpha})    =
e^{-ig\mathbf{k}\cdot {\bf L}_{\alpha}}
U_{g}^{l_{\alpha};l^{\prime}_{\alpha^{\prime}}}\phi_{ \alpha^{\prime} l_{\alpha^{\prime}}^{\prime}}(  g\mathbf{k}).
\end{align}
\end{widetext}
Returning to the \textquotedblleft orbital\textquotedblright\ notation of the
main text, we combine $\alpha$ and $l_{\alpha}$ into a single large vector with
index $l\equiv (  \alpha,l_{\alpha}) $. The phase $e^{-ig\mathbf{k}\cdot{\bf L}_{\alpha}}$ can then be represented by a
diagonal unitary matrix $\mathbf{U}_{g\mathbf{k},{\bf L}}^{^{\prime}}$ that
depends on $g$ and also on the $\alpha$ part of the index $l$. Similarly, using the
new indices, $U_{g}^{l_{\alpha};l^{\prime}_{\alpha^{\prime}}}$ can be rewritten
so that all indices can be taken into account at once, leading, in vector
notation, to
\begin{equation}
\left\{  g|{\bm \tau}\right\}  \phi(  \mathbf{k})  =
%e^{-i\mathbf{k}\cdot{\bm \tau}}
\mathbf{U}_{g\mathbf{k,L}}^{\prime}\mathbf{U}_{g}\phi(  g\mathbf{k})  .
\end{equation}

\section{Symmetry considerations for a minimal tight-binding Hamiltonian}\label{TBHamiltonian}
In this section we further clarify the above symmetry concepts using a minimal Hamiltonian where
%. Consider a tri-layer Fe-As plan in which As ions are puckering above and bellow of the Fe plan as it is shown in \fref{fig0}. Here, 
we consider only Fe-$d_{x^2-y^2}$ and  As-$p_z$ orbitals. The unit cell includes two Fe ions and two As ions, hence the Hamiltonian is a $4 \times 4$ matrix.  For the sake of simplicity, we consider a quasi-$2D$ system while in reality there is a finite dispersion in the $z$ direction. 
The unit cell vectors are ${\bf a}_1=(a,0,0)$ and ${\bf a}_2=(0,a,0)$ and ion positions are ${\bf r}_{A-{\rm Fe}}=(-a/4,-a/4,0)$, ${\bf r}_{A-{\rm As}}=(-a/4,a/4,-h)$, ${\bf r}_{B-{\rm Fe}}=(a/4,a/4,0)$, and ${\bf r}_{B-{\rm As}}=(a/4,-a/4,h)$, where $A/B$ label sublattices and the origin is on the inversion center. 

\subsection{The minimal Hamiltonian and its symmetry properties}
\begin{figure}
\centering{
	\includegraphics[width=0.6\linewidth,clip=]{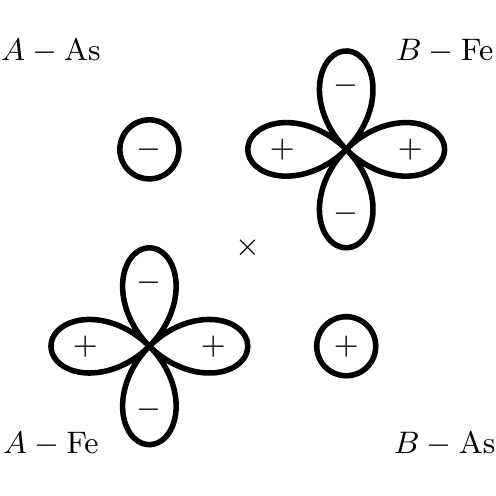}}
	\caption{ Schematics of a FeAs tri-layer unit cell with Fe-$d_{x^2-y^2}$ and As-$p_z$.   Note that for As-$p_z$ only the sign of orbital's lobe near to the Fe-Fe plan is shown. The hopping matrix elements between Fe-$d_{x^2-y^2}$ and As-$p_z$ change sign by changing Fe ion. The cross shows the inversion center.
	}\label{fig0}
\end{figure}

The  Hamiltonian is given by ${\bf {\mathcal H}}=\sum_{{\bf k}}{\bm \phi}^{\dagger}_{{\bf k}}{\bf H}_{\bf k} {\bm \phi}_{{\bf k}}$  where  
the basis set is 
\begin{equation}
{\bm \phi}_{{\bf k}}\equiv\{\phi_{\rm A-Fe}({\bf k}),\phi_{\rm A-As}({\bf k}),\phi_{\rm B-Fe}({\bf k}),\phi_{\rm B-As}({\bf k}) \}^T.  
\end{equation}
With only nearest-neighbor hopping, the Hamiltonian, \eref{Ham1}, in the standard gauge is given by 
\begin{equation}
%{\bf H}_{\bf k}=
\begin{bmatrix}  
\epsilon_d& t(1+z_{y{\bf k}}^*) & 0& t(1+z_{x{\bf k}}^*)\\ 
t(1+z_{y{\bf k}}) & \epsilon_p & -t(1+z_{x{\bf k}}^*)&0 \\
0&-t(1+z_{x{\bf k}}) &\epsilon_d & -t(1+z_{y{\bf k}})\\
 t(1+z_{x{\bf k}})&0&-t(1+z_{y{\bf k}}^*) & \epsilon_p
 \end{bmatrix}, \label{Ham2}
\end{equation}
where $t$ denote the nearest-neighbor hopping amplitude between  As-$p_z$ orbitals and Fe-$d_{x^2-y^2}$.  denote On-site energies are $\epsilon_d$, $\epsilon_p$ and we defined 
\begin{equation} 
z_{x/y}({\bf k}) \equiv \exp(-ik_{x/y}a).  
\end{equation}
From \fref{fig0} one can see that the hopping amplitude between Fe and As has  different sign for Fe-$A$ and Fe-$B$. 
%In real system there is an additional ${\bf H}_{\perp, \bf k}$ which we discuss its effect perturbatively later.  
The Hamiltonian in the alternative gauge, ${\bf H}^{\prime}_{\bf k}$, is obtained from \eref{Ham2} by substituting $t$ with $t\exp[-i{\bf k}\cdot ({\bf r}_{1}-{\bf r}_2)]$ where ${\bf r}_{1/2}$ denotes the orbital position. 

The Hamiltonian ${\bf H}_{\bf k}$  is trivially invariant under identity. Under inversion symmetry $\{I|000\}$, the wave vectors transform as $(k_x,k_y,k_z) \rightarrow (-k_x,-k_y,-k_z)$. The transformation matrix for the orbitals is given by
\begin{equation}
{\bf U}_{I}=
\begin{bmatrix}  
0&0&1&0\\
0&0&0&-1\\
1&0&0&0\\
0&-1&0&0
 \end{bmatrix}, \label{Transfer1}
\end{equation}
which gives  
${\bf U}^{\dagger}_{I}{\bf H}_{\bf k}{\bf U}_{I}={\bf H}_{I \bf k}={\bf H}(-k_x,-k_y,-k_z)$. Note that the basis set transforms accordingly, ${\bm \phi}_{\bf k} \rightarrow {\bf U}_{I}{\bm \phi}_{I{\bf k}}$.  

Another symmorphic generator of the group is $\{C_{2a}|000\}$. Under $\{C_{2a}|000\}$ the wave vector transforms as $(k_x,k_y,k_z) \rightarrow (k_y,k_x,-k_z)$ and the orbital space transformation matrix is 
\begin{equation}
{\bf U}_{C_{2a}}=
\begin{bmatrix}  
-1&0&0&0\\
0&0&0&-1\\
0&0&-1&0\\
0&-1&0&0
 \end{bmatrix}. \label{Transfer2}
\end{equation}
Again the transformed Hamiltonian is 
${\bf U}^{\dagger}_{C_{2a}}{\bf H}_{\bf k}{\bf U}_{C_{2a}}={\bf H}_{C_{2a} \bf k}={\bf H}(k_y,k_x,-k_z)$. 

The last generator of the group is a non-symmorphic symmetry operation, i.e., $\{C_{2y}|0\frac{1}{2}0\}$. Under this symmetry operation $(k_x,k_y,k_z) \rightarrow (-k_x,k_y,-k_z)$. If one does not consider the extra phases induced by the half-integer translation, then in the truncated orbital space considered in this subsection, i.e., Fe-$d_{x^2-y^2}$ and As-$p_z$, the orbital-space transformation matrix is the same as for inversion, \eref{Transfer1}. Since the wave-vector transformation is different from that of inversion symmetry, one can immediately conclude  that  ${\bf U}^{\dagger}_{C_{2y}}{\bf H}_{\bf k}{\bf U}_{C_{2y}}$ is not equal to ${\bf H}_{C_{2y} \bf k}={\bf H}(-k_x,k_y,-k_z)$. Due to extra phases, in this case the correct orbital-space transformation matrix is
\begin{equation}
{\bf U}_{\{C_{2y}|0\frac{1}{2}0\}}({\bf k})=
\begin{bmatrix}  
0&0&1&0\\
0&0&0&-z_{y,{\bf k}}\\
z_{y,{\bf k}}&0&0&0\\
0&-1&0&0
 \end{bmatrix}, \label{Transfer3}
\end{equation}
where the extra phases appear for As-A and Fe-B that leave the unit cell under this symmetry operation. 
Therefore, 
${\bf H}_{C_{2y} \bf k}$ is equal to ${\bf U}^{ \dagger}_{\{C_{2y}|0\frac{1}{2}0\}}({\bf k}){\bf H}_{\bf k}{\bf U}_{\{C_{2y}|0\frac{1}{2}0\}}({\bf k})$ where, explicitly
\begin{align}
{\bf U}&^{ \dagger}_{\{C_{2y}|0\frac{1}{2}0\}}({\bf k}){\bf H}_{\bf k}{\bf U}_{\{C_{2y}|0\frac{1}{2}0\}}({\bf k})=\nonumber\\
&\begin{bmatrix}  
H_{33}&-z^*_{y,{\bf k}}H_{34}&z^*_{y,{\bf k}}H_{31}&-H_{32}\\
-z_{y,{\bf k}}H_{43}&H_{44}&-H_{41}&z_{y,{\bf k}}H_{42}\\
z_{y,{\bf k}}H_{13}&-H_{14}&H_{11}&-z_{y,{\bf k}}H_{12}\\
-H_{23}&z^*_{y,{\bf k}}H_{24}&-z^*_{y,{\bf k}}H_{21}&H_{22}
 \end{bmatrix}.\label{UHU}
\end{align}
Here, $H_{ij}$'s are matrix elements of ${\bf H}_{\bf k}$ given in \eref{Ham2}. For example, consider the $(1,2)$ component. It is given by $-z^*_{y,{\bf k}}H_{34}$ where, using $H_{34}$ from \eref{Ham2}, it is $tz^*_{y,{\bf k}}(1+z_{y{\bf k}})=t(1+z^*_{y{\bf k}})= H_{12}(-k_x,k_y,-k_z)$.
The above equation is also valid in the alternative gauge if one replaces the matrix element of ${\bf H}_{\bf k}$ with ${\bf H}^{\prime}_{\bf k}$ and employs the fact that 
the hopping matrix elements $t\exp[-i{\bf k}\cdot ({\bf r}_{1}-{\bf r}_2)]$ can be rewritten as $t\exp[iC_{2y}{\bf k}\cdot C_{2y}({\bf r}_{1}-{\bf r}_2)]$.  
We now have the transformation matrices for all of the generators of the group.

Finally, we are also interested in the four-fold symmetry that can obtained as $\{C^+_{4z}|0\frac{1}{2}0\}$. Under this symmetry the wave vector changes as follows,  $(k_x,k_y,k_z)\rightarrow(-k_y,k_x,k_z)$ and the orbital space transforms as
\begin{equation}
{\bf U}_{\{C^+_{4z}|0\frac{1}{2}0\}}=
\begin{bmatrix}  
0&0&-1&0\\
0&1&0&0\\
-z_{x,{\bf k}}&0&0&0\\
0&0&0&z_{x,{\bf k}}
 \end{bmatrix}. \label{Transfer4}
\end{equation}
For $\{C^+_{4z}|0\frac{1}{2}0\}$  the half-translation is in the $y$ direction, so ${\bf L}={\bf a}_2$ for As and Fe ions on the $B$ sublattice and the argument of the exponential is $-ig{\bf k}\cdot {\bf L}$, which gives $z_{x,{\bf k}}$ for the extra phases.

\subsection{Eigensystem of the minimal Hamiltonian}
Now, let us consider the symmetry properties of the eigenstates. In this section we focus on four-fold symmetry in the $k_z=0$ plane. The eigenenergies of the Hamiltonian \eref{Ham1} are~\cite{Note1}
%~\footnote{It is straightforward to show that the \eref{Ham1} satisfies the following identity
%$\big([{\bf H}_{\bf k}/t-(\epsilon_d+\epsilon_p)/2t{\bm 1}]^2-[(\epsilon_d-\epsilon_p)^2/4t^2+\lambda^2_{x{\bf k}}+\lambda^2_{y{\bf k}}]{\bm 1}\big)^2=4\lambda^2_{x{\bf k}}\lambda^2_{y{\bf k}}{\bm 1}$.}
%
\begin{align}
E_{\pm\pm,{\bf k}} = &\frac{(\epsilon_d+\epsilon_p)}{2}\pm \sqrt{\Delta^2+(\lambda_{x{\bf k}}\pm\lambda_{y{\bf k}})^2}\nonumber\\
\equiv&\frac{(\epsilon_d+\epsilon_p)}{2}\pm E^{\prime}_{\pm\bf k},
\end{align}
where $\Delta\equiv (\epsilon_d-\epsilon_p)/2$ and $\lambda^2_{x/y{\bf k}} \equiv t^2(1+z_{x/y{\bf k}})(1+z^*_{x/y{\bf k}})=t^2[(1+\cos k_{x/y}a)^2+\sin^2 k_{x/y}a]=4t^2\cos^2 (k_{x/y}a/2)$. The eigenenergies have four-fold symmetry, i.e. $E_{\pm\pm,{\bf k}}$ ($E^{\prime}_{\pm \bf k}$) are invariant under $(k_x,k_y)\rightarrow(k_y,-k_x)$ since under this transformation we have $(\lambda^2_{x{\bf k}},\lambda^2_{y{\bf k}})\rightarrow (\lambda^2_{y{\bf k}},\lambda^2_{x{\bf k}})$. As expected on general principles for non-symmorphic space groups, the bands are degenerate on the BZ boundaries where $k_x=\pm \pi/a$ or $k_y=\pm \pi/a$ lead to vanishing $\lambda_x$ or $\lambda_y$.

By diagonalizing \eref{Ham1}, the expansion coefficients $a^l_{n{\bf k}}=\langle \phi_{l{\bf k}}|\psi_{n{\bf k}}\rangle$ for $E_{\bf k} = (\epsilon_d+\epsilon_p)/2-(+) E^{\prime}_{\pm\bf k}$ at a ${\bf k}$-point away from the diagonals are obtained as
\begin{align}
|a_{-(+) \pm,{\bf k}}\rangle\propto \big\{-&\frac{(\lambda_{x{\bf k}}\pm \lambda_{y{\bf k}})}{(\Delta+(-) E^{\prime }_{\pm \bf k})}\exp(i{\bf k}\cdot {\bf r}_{A-Fe}),\nonumber\\
\pm&\exp(i{\bf k}\cdot {\bf r}_{A-As}),\nonumber \\
&\frac{(\lambda_{y{\bf k}}\pm \lambda_{x{\bf k}})}{(\Delta+(-) E^{\prime }_{\pm \bf k})}\exp(i{\bf k}\cdot {\bf r}_{B-Fe}),\nonumber \\
&\exp(i{\bf k}\cdot {\bf r}_{B-As}) \big\}^T.\label{Coeff}
\end{align}
For the $\bf k$-points  on the primary diagonal ($k_x=k_y$) and secondary diagonal ($k_x=-k_y$), we have the equalities $\lambda_{x{\bf k}}=\lambda_{y{\bf k}}$ and  $E^{\prime}_{-\bf k} =\Delta$. The eigenstate with energy $E_{+-,\bf k} = (\epsilon_d+\epsilon_p)/2+ E^{\prime}_{-,\bf k}$ for these ${\bf k}$-points is
\begin{align}
|a_{+ -,{\bf k}}\rangle\propto \big\{\exp(i{\bf k}\cdot {\bf r}_{A-Fe}),
0,
\exp(i{\bf k}\cdot {\bf r}_{B-Fe}),
0 \big\}^T.\label{Coeff2}
\end{align}
Finally, for the $M=(\pm \pi/a,\pm \pi/a)$ the eigenvector of the $E_{\bf k} = (\epsilon_d+\epsilon_p)/2+ E^{\prime}_{+,\bf k}$  may be chosen as 
\begin{align}
|a_{- -,{\bf k}}\rangle\propto \big\{i\frac{\sqrt{2}}{2}\frac{k_x}{|k_x|},
0,
i\frac{\sqrt{2}}{2}\frac{k_y}{|k_y|},
0 \big\}^T.\label{Coeff3}
\end{align}
The norms of the Bloch function, \eref{Coeff}, are equal to
\begin{align}
\langle a_{-(+) \pm,{\bf k}} &|a_{-(+) \pm,{\bf k}}\rangle = 2 +\nonumber\\
&\frac{(\lambda_{x{\bf k}}\pm \lambda_{y{\bf k}})^2+(\lambda_{y{\bf k}}\pm \lambda_{x{\bf k}})^2}{(\Delta -(+)  E^{\prime }_{\pm \bf k})^2},\label{norm}
\end{align}
which has four-fold symmetry.  
%In the alternative gauge, the $a^l_{n{\bf k}}$ are replaced by $b^l_{n{\bf k}}=\exp(-i{\bf k}\cdot {\bf r}_l)a^l_{n{\bf k}}$ and $\phi_{l{\bf k}}({\bf r})$ by $\phi^{\prime}_{l{\bf k}}({\bf r})=\exp(i{\bf k}\cdot {\bf r}_l)\phi_{l{\bf k}}({\bf r})$.

The Bloch states  are gauge independent up to an overall phase.  So, it does not matter whether we use the natural gauge or the alternative gauge.
%The periodic part of the Bloch functions are given by $u_{n{\bf k}}({\bf r})=\exp(i{\bf k}\cdot {\bf r})  \psi_{n{\bf k}}({\bf r})$. Since the Bloch functions are gauge independent, $u_{n{\bf k}}({\bf r})$ inherent this property. 
Even though eigenenergies are invariant under four-fold symmetry, it is clear that the Bloch functions are not. However, the Bloch functions are auxiliary quantities and are not observable. On the other hand, an observable quantity such as the electron density $n({\bf r})$ is given by $\sum_{n{\bf k}} f(\epsilon_{n{\bf k}})|\psi_{n,{\bf k}}({\bf r})|^2 $ where $f(\epsilon_{n{\bf k}})$ %$f(\epsilon_{n{\bf k}})=1/[\exp({\beta(\epsilon_{n{\bf k}}-\mu)})+1]$ 
is the Fermi-Dirac distribution function.  It has four-fold symmetry because the norm of the expansion coefficients, \eref{norm}, has this symmetry. 

\section{Symmetry of the in-plane response function of the minimal Hamiltonian}\label{Sec:SymmInPlane}

Here we show explicitely that despite the fact that the generalized susceptibility \eref{Chi0} obtained from the minimal model Hamiltonian does not have $C_{4z}$ symmetry for $q_z=0$, the observable susceptibility at $q_z=0$ does. We need to find how the observable in-plane response function transforms under $\{C_{4z}|0\frac{1}{2}0\}$ by inspecting the symmetry properties of \eref{ChargeSus}, which includes the oscillator matrix elements. In this section, we assume that the applied field wave-vector lies within the BZ and set ${\bf G}, {\bf G}' = {\bm 0}$. In order to find how the in-plane $\chi({\bf q},\nu)$  with $q_z=0$ transforms to $\chi(C_{4z}{\bf q},\nu)$ under $\{C_{4z}|0\frac{1}{2}0\}$ symmetry, we need to transform each term in $\chi({\bf q},\nu)$ to the corresponding term with $({\bf k},{\bf q}) \rightarrow C_{4z}({\bf k},{\bf q})$,  $A-Fe \leftrightarrow B-Fe$ and invariant $As$ ions. Note that $\epsilon_{n,{\bf k}}=\epsilon_{n,C_{4z}{\bf k}}$, so we focus on the transformation properties of the expansion coefficients and the oscillator matrix elements in \eref{ChargeSus} and \eref{Chi0}.

For example,  consider the following contribution  which includes inter-site oscillator matrix elements between Fe sites and As 
\begin{align}
a&^{A-As*}_{n,{\bf k}-{\bf q}}a^{A-Fe}_{m,{\bf k}}a^{B-As*}_{m,{\bf k}}a^{B-Fe}_{n,{\bf k}-{\bf q}}\nonumber\\&\times
O_{{\bf k}}^{A-As,A-Fe}({\bf q})O_{{\bf k}}^{B-As,B-Fe}(-{\bf q}).\label{SB}
\end{align}
We need to sum over all $k_z$ to compute the response functions, but since $k_z$ is invariant under $C_{4z}$, we present for simplicity the analysis for the $k_z=0$ case only. By comparing $|a_{n,{\bf k}}\rangle$ and $|a_{n,C_{4z}{\bf k}}\rangle$, where $C_{4z}(k_x,k_y)\rightarrow (-k_y,k_x)$, one can verify that the expansion coefficients of a general in-plane $\bf k$-point, \eref{Coeff},  satisfy  
\begin{align}
a^{A-As}_{n,{\bf k}} &= \exp(-ik_xa/2)a^{A-As}_{n,C_{4z}{\bf k}},\nonumber\\
a^{A-Fe}_{n,{\bf k}} &= -\exp(-ik_xa/2)a^{B-Fe}_{n,C_{4z}{\bf k}},\nonumber\\
a^{B-As}_{n,{\bf k}} &= \exp(ik_xa/2)a^{B-As}_{n,C_{4z}{\bf k}},\nonumber\\
a^{B-Fe}_{n,{\bf k}} &= -\exp(ik_xa/2)a^{A-Fe}_{n,C_{4z}{\bf k}}.\label{Coef}
\end{align}
The negative signs in the second and fourth equations in \eref{Coef} do not exist for momenta on the diagonals and the band corresponding to \eref{Coeff2}. However, in our simple case, \eref{SB} is identically zero for these momenta and band index because the expansion coefficients on As sites are zero, hence our discussion is valid in general.  However, as we proceed to show momentarily, the existence of these negative signs is essential to recover the symmetry. Therefore, any perturbation that causes a non-zero correction to the As expansion coefficients in \eref{Coef} will break four-fold symmetry.

Under $\{C_{4z}|0\frac{1}{2}0\}$ the oscillator matrix elements, \eref{OM}, transform as (see \eref{Transfer4} and its discussion)
\begin{align}
\{C_{4z}|0\frac{1}{2}0\}O_{{\bf k}}^{A-As,A-Fe}({\bf q})\rightarrow& \nonumber\\
-O_{C_{4z}{\bf k}}^{A-As,B-Fe}&(C_{4z}{\bf q}),\nonumber\\
\{C_{4z}|0\frac{1}{2}0\}O_{{\bf k}}^{B-As,B-Fe}(-{\bf q})\rightarrow\nonumber\\
-e^{iq_xa}
O_{C_{4z}{\bf k}}^{B-As,A-Fe}&(-C_{4z}{\bf q}),\label{Osci2}
\end{align}
where we used passive rotation and the identity $\phi^{A/B-As}_{\bf k}(C_{4z}^{-1}{\bf r}-{\bm \tau})=\phi^{A/B-As}_{C_{4z}{\bf k}}({\bf r})$ and $\phi^{A/B-Fe}_{\bf k}(C_{4z}^{-1}{\bf r}-{\bm \tau})=-\phi^{B/A-Fe}_{C_{4z}{\bf k}}({\bf r})$.  The negative sign in the latter identity comes from the $d_{x^2-y^2}$ orbital symmetry.

By replacing \eref{Coef} for the expansion coefficients and \eref{Osci2} for the oscillator matrix element in \eref{SB}, one can see that the signs and  the extra phases  cancel out and  \eref{SB} transforms as
\begin{align}
a^{A-As*}_{n,C_{4z}({\bf k}-{\bf q})}&a^{B-Fe}_{m,C_{4z}{\bf k}}a^{B-As*}_{m,C_{4z}{\bf k}}a^{A-Fe}_{n,C_{4z}({\bf k}-{\bf q})}\times\nonumber\\
O_{C_{4z}{\bf k}}^{B-Fe,A-As}&(C_{4z}{\bf q})O_{C_{4z}{\bf k}}^{B-As,A-Fe}(-C_{4z}{\bf q}).
\end{align}
Hence,  \eref{SB} maps into the corresponding term in $\chi(C_{4z}{\bf q},\nu)$ under four-fold symmetry. Clearly, the symmetry properties of both the generalized susceptibility, \eref{Chi0} or \eref{eq:bareSusRPA}, and the oscillator matrix elements should be considered to obtain an in-plane four-fold response function. On the other hand, the in-plane generalized susceptibility by itself is not invariant under four-fold symmetry.

%\bibliography{manuscript}


\end{document}